\documentclass[letterpaper]{article} 
\usepackage{aaai24}  
\nocopyright
\usepackage{times}  
\usepackage{helvet}  
\usepackage{courier}  
\usepackage[hyphens]{url}  
\usepackage{graphicx} 
\usepackage{natbib}  
\usepackage{caption} 
\usepackage{algorithm}
\usepackage{algorithmic}
\usepackage{amsmath}
\usepackage{amsthm}
\usepackage{amssymb}
\usepackage{commath}
\usepackage{enumitem}
\usepackage{empheq}
\usepackage{booktabs}
\usepackage{newfloat}
\usepackage{listings}
\DeclareCaptionStyle{ruled}{labelfont=normalfont,labelsep=colon,strut=off} 
\floatstyle{ruled}
\newfloat{listing}{tb}{lst}{}
\floatname{listing}{Listing}

\title{Search Me Knot, Render Me Knot: Embedding Search and Differentiable Rendering of Knots in 3D}
\author {
    Aalok Gangopadhyay\textsuperscript{\rm 1},
    Paras Gupta\textsuperscript{\rm 1},
    Tarun Sharma\textsuperscript{\rm 1}
    Prajwal Singh\textsuperscript{\rm 1}
    Shanmuganathan Raman\textsuperscript{\rm 1}
}
\affiliations {
    \textsuperscript{\rm 1}Indian Institute of Technology, Gandhinagar, India\\
    \{aalok, gupta.paras, sharma\_tarun, singh\_prajwal, shanmuga\}@iitgn.ac.in
}

\begin{document}

\maketitle

\begin{abstract}
We introduce the problem of knot-based inverse perceptual art. Given multiple target images and their corresponding viewing configurations, the objective is to find a 3D knot-based tubular structure whose appearance resembles the target images when viewed from the specified viewing configurations. To solve this problem, we first design a differentiable rendering algorithm for rendering tubular knots embedded in 3D for arbitrary perspective camera configurations. Utilizing this differentiable rendering algorithm, we search over the space of knot configurations to find the ideal knot embedding. We represent the knot embeddings via homeomorphisms of the desired template knot, where the homeomorphisms are parametrized by the weights of an invertible neural network. Our approach is fully differentiable, making it possible to find the ideal 3D tubular structure for the desired perceptual art using gradient-based optimization. We propose several loss functions that impose additional physical constraints, enforcing that the tube is free of self-intersection, lies within a predefined region in space, satisfies the physical bending limits of the tube material and the material cost is within a specified budget. We demonstrate through results that our knot representation is highly expressive and gives impressive results even for challenging target images in both single view as well as multiple view constraints. Through extensive ablation study we show that each of the proposed loss function is effective in ensuring physical realizability. We construct a real world 3D-printed object to demonstrate the practical utility of our approach. To the best of our knowledge, we are the first to propose a fully differentiable optimization framework for knot-based inverse perceptual art.
\end{abstract}


\section{Introduction}
\label{sec:introduction}

\begin{figure}[!t]
\centering
    \begin{minipage}[b]{0.49\linewidth}
        \centerline{\includegraphics[width=\linewidth]{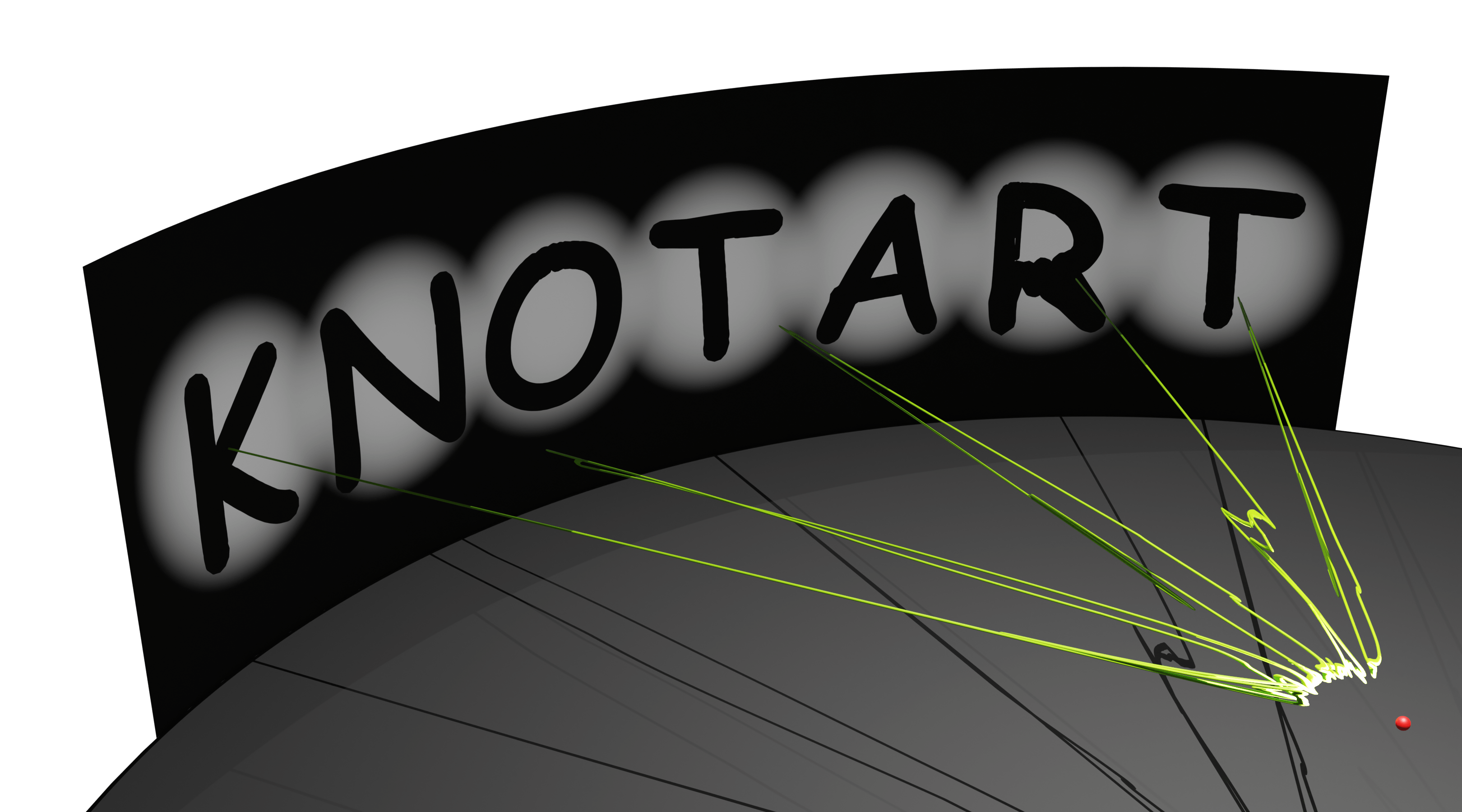}}
        \centerline{\includegraphics[width=\linewidth]{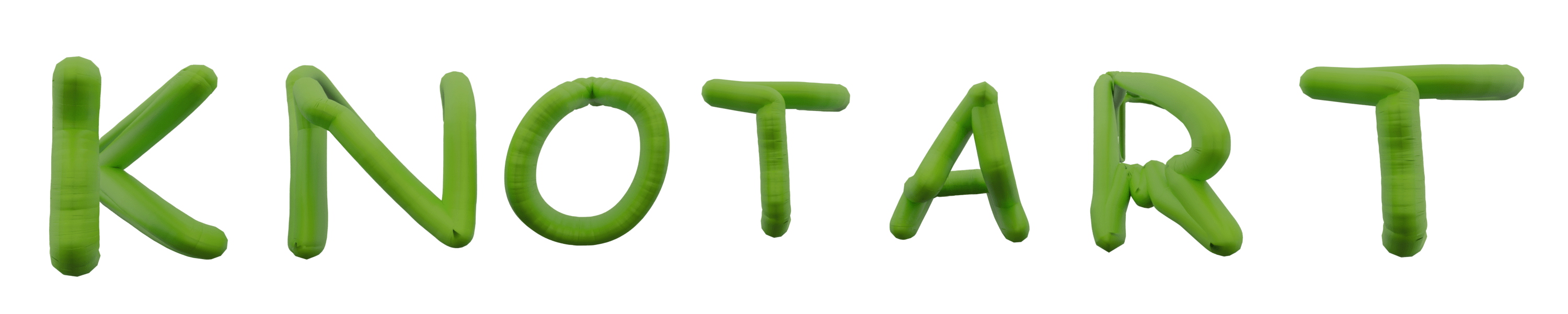}}
        \centerline{(a) Alphabetical Images}\medskip
   \end{minipage}
   \begin{minipage}[b]{0.49\linewidth}
        \centering
        \centerline{\includegraphics[width=\linewidth]{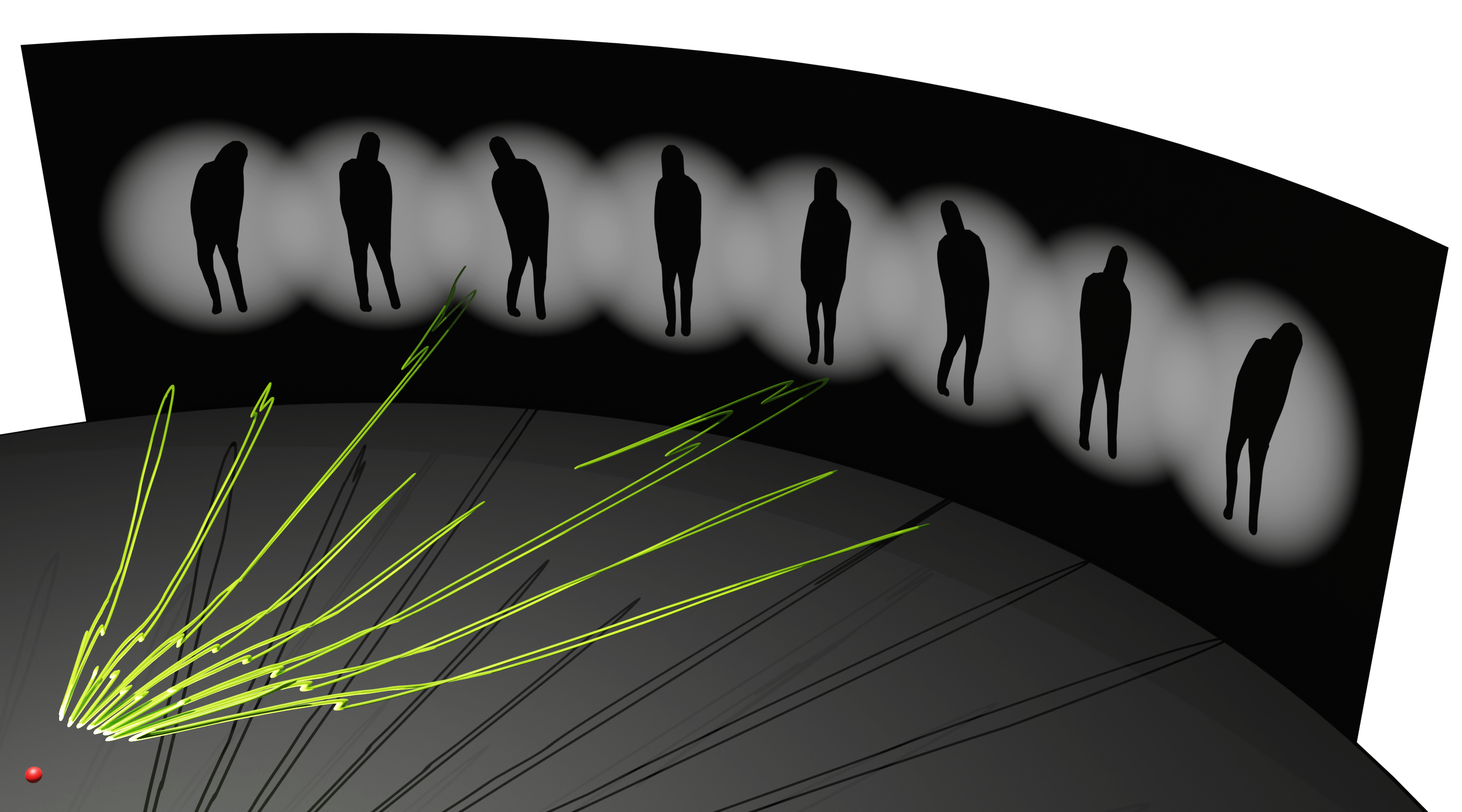}}
        \centerline{\includegraphics[width=\linewidth]{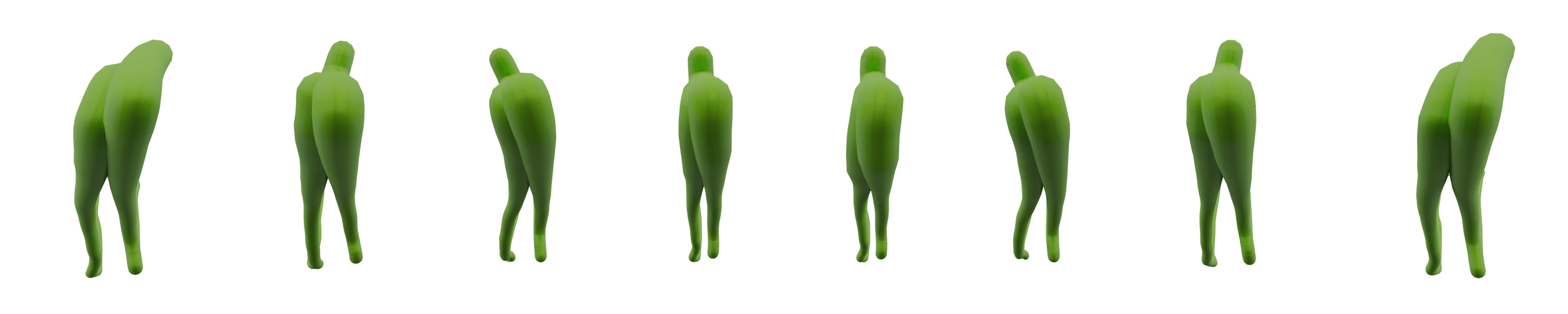}}
        \centerline{(b) Dancing Person}\medskip
   \end{minipage}

    \captionof{figure}{Results obtained by our approach for (a) alphabetical images and (b) dancing person image sequence. The red dot indicates the viewer location as well as the light location.
    Our method searches for the embedding of the tubular-knots (shown in green), so that the projected images resemble the target images. 
    In both the cases, the top row image depicts the shadow cast on a nearby gallery wall by a light source placed at the red dot and the bottom row depicts the image a viewer will perceive if viewing from the red dot. }
    \label{fig:Teaser}
\end{figure}

Suppose that an artist wishes to create a perceptual artwork in three-dimensional space by creating a structure that appears meaningful only when seen from specific viewpoints in the scene and appears arbitrary or meaningless from other viewpoints. 
With such a goal set, the artist has in mind the target visual images that should be perceived from specific viewpoints. Figuring out the 3D structure that would resemble the target images is an inverse process, which is very challenging and requires human ingenuity. Consider another scenario where light sources are placed at specific locations, such that when the light gets cast onto the 3D structure, a shadow is formed on a nearby wall. The artist would desire the shadow formed to match the target image in this scenario.
In the first case, the viewer is at a specific location, and in the second, the light source is at a specific location. Despite the differences, these scenarios involve the same underlying principle of perspective projection, where the target 2D image is the known parameter, and the 3D art structure is the unknown parameter. As an example, consider Fig. \ref{fig:Teaser}, where given a specific viewpoint (red dot), the top image depicts the shadow cast of 3D structures (green tube) on a nearby wall cast by a light placed at the viewpoint and the bottom image depicts the image perceived by a viewer situated at the viewpoint.

Drawing inspiration from the wire-sculpture shadow art of Larry Kagan  \cite{larry-kagan,BlauweissMedia_2023}, here we consider a specific form of perceptual art, where the 3D structure is made of tube-based shapes like wires, ropes or strings. A tube here refers to the volume sweep of a sphere along a curve embedded in 3D space. For the curves, we specifically consider knots or loops, which are closed curves in 3D. Given a desired target silhouette image and the tube thickness, we search for the embedding or the configuration of the knot in 3D so that the knot-based tube, when viewed from a given camera viewpoint, resembles the target image. Moreover, while creating such a 3D structure, the artist would also need to ensure that the structure is physically realizable; i.\,e., it should be free of self-intersections, and given the material that will be used for the tube, the structure should respect the physical constraints, for instance, the maximum allowed bending. Since such a structure is to be realised in the real world, the artist might want the tubular structure to fit inside a predefined region in space in the scene. Also, taking the material cost into consideration, the artist might want to minimize the usage of the material, which in this case would be directly proportional to the length of the knot. 

We have proposed an end-to-end differentiable optimization framework to guide an artist in finding a 3D structure corresponding to the desired target image(s). We search for the 3D art structure by searching for knot embeddings in 3D, where we define the search space through a parametrized family of homeomorphisms represented using an Invertible Neural Network (INN). 
We have designed a differentiable rendering algorithm that, given a knot embedding, the thickness of the tube, and the information about the perspective viewing camera placed at a specific viewpoint, generates a silhouette rendering of the tube that would be perceived from that specific viewpoint.
Our approach has the theoretical guarantee of the knot embedding being free of self-intersection. To avoid self-intersections in the tube, we have developed an appropriate loss function. We also penalize high curvature regions, thereby restricting the tube bending. Material cost constraint is satisfied by adding a penalty whenever the knot length exceeds the allowed limit, whereas space constraint is satisfied by penalizing segments of the tube that move outside the predefined region in the scene.

The following are the major contributions of this work:

\begin{enumerate}
    \item We develop an end-to-end differentiable framework for finding a 3D art structure based on tubular knots, which is perceived as the desired target image upon viewing from a particular viewpoint. Our approach also works in the multi-view setting, where we have multiple target images with corresponding viewpoint configurations.
    \item We design an efficient differentiable rendering algorithm that, given an arbitrary camera configuration, renders the silhouette image of a tube of desired thickness along a given knot embedding in 3D, as viewed from the camera.  
    \item We propose a differentiable neural representation of knot embeddings through Invertible Neural Networks (INN), facilitating effective knot search using gradient-based optimization.
    \item We propose several loss functions to ensure the 3D art structure is physically realizable.
\end{enumerate}

To the best of our knowledge, we are the first to formalize the problem of knot-based inverse perceptual art in the optimization setting and to propose a fully differentiable inverse rendering-based solution for the same.
\section{Related Work}
\label{sec:RelatedWork}

\textbf{Shadow Art.}
The inverse process of reconstructing or learning 3D structures that cast target shadows under specific lighting conditions has been previously explored. \cite{mitra2009shadow} deform the input target images to find a consistent shadow hull while minimizing the induced distortions, utilising volumetric representation. \cite{sadekar2022shadow} use a differentiable rendering optimization framework for finding the 3D structure, using mesh representation. Our work in contrast focuses on knot-based tubular 3D structures and utilizes neural representations for representing the knot. One of the drawbacks in the previous works is that the predicted 3D structure is not guaranteed to be physically realisable, limiting its practical applicability. Our work in contrast puts significant emphasis on physical realizability making it relevant for real-world applications.


\begin{figure}[!t]
    \begin{minipage}[b]{0.32\linewidth}
        \centering
        \includegraphics[width=\linewidth]{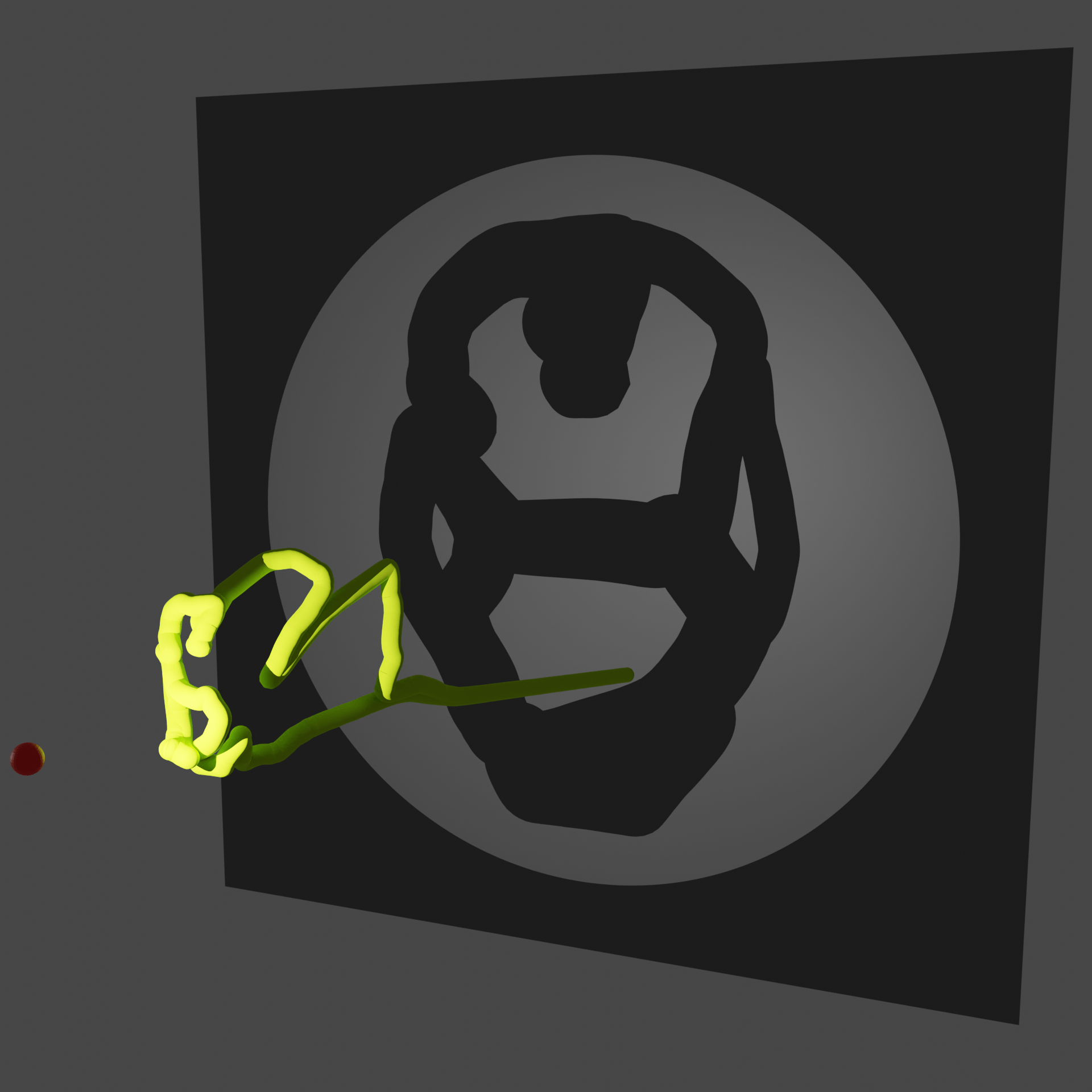}\\
        \includegraphics[width=\linewidth]{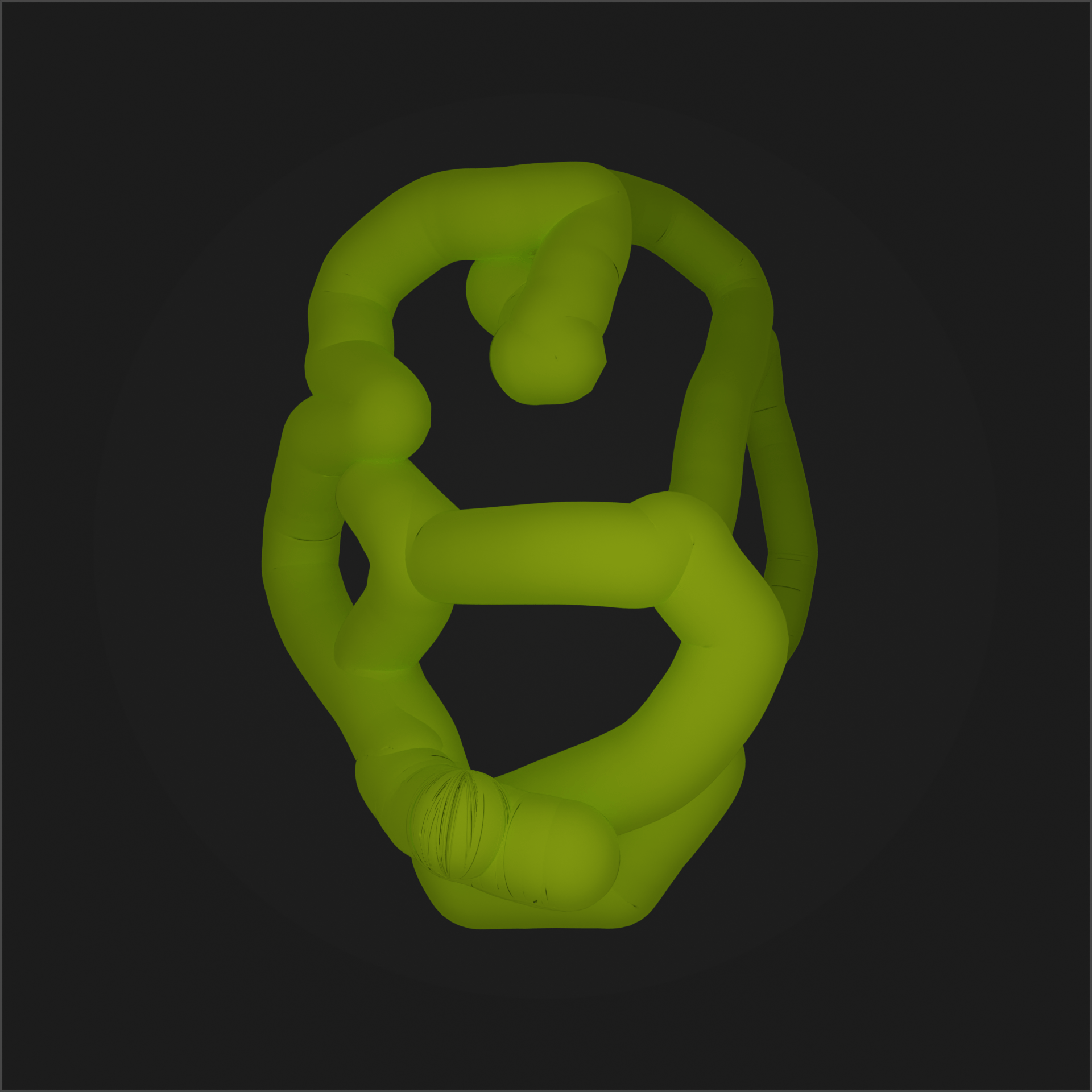}\\
        (a) Single Viewer\medskip
   \end{minipage}
   \begin{minipage}[b]{0.64\linewidth}
        \centering
        \includegraphics[width=\linewidth]{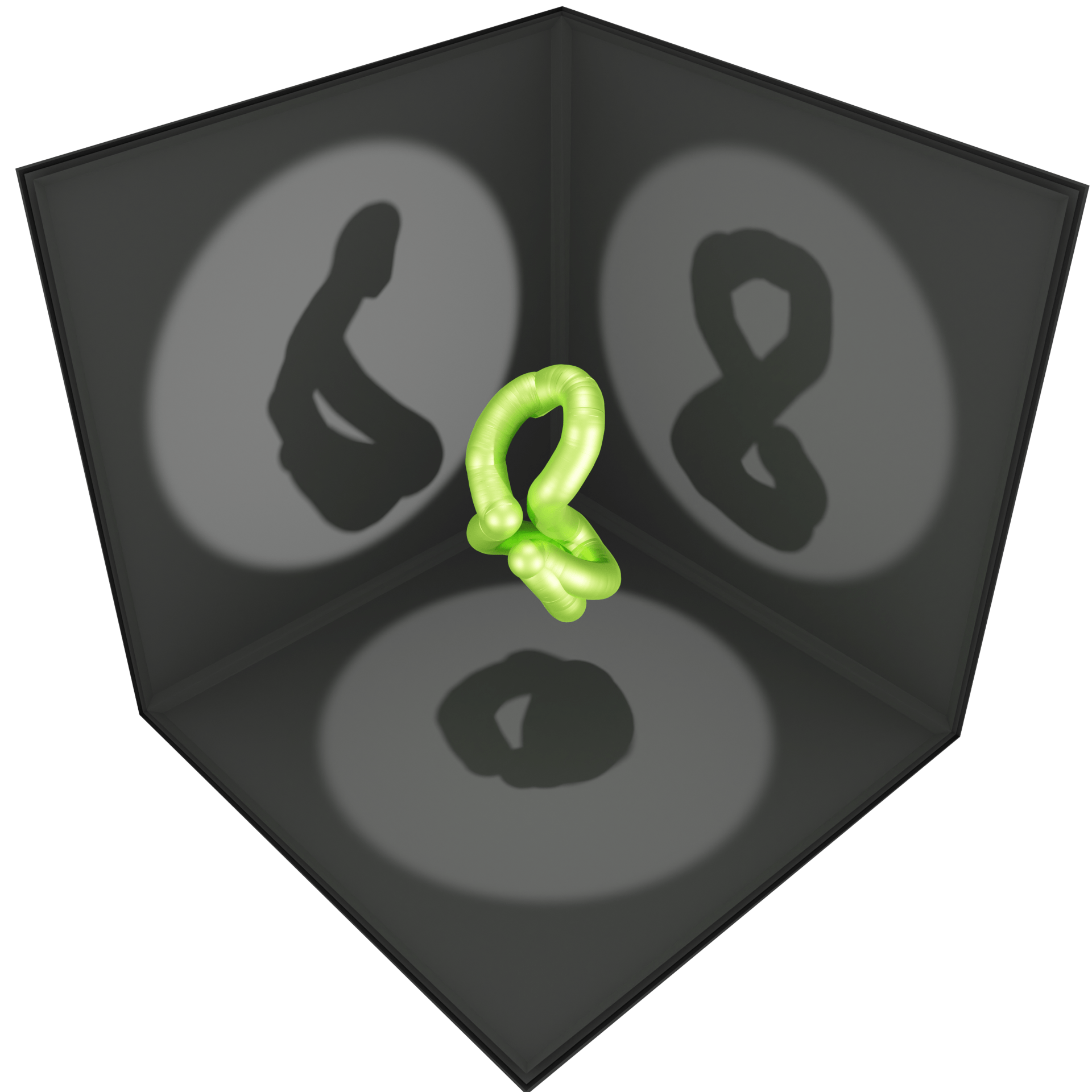}\\
        (b) Multiple Viewer\medskip
   \end{minipage}
    \captionof{figure}{(a) Result on Iron Man Image. The top image shows the shadow cast on the wall, whereas the bottom image shows the view from the red dot.
    (b) Given the orthographic projections of Viviani's curve as multiple target images, our method learns a knot embedding whose perspective projections approximate the target images when viewed from the respective viewing locations.}
    \label{fig:ironman_viviani}
\end{figure}%

\begin{figure*}[!htb]
\centering
\centerline{\includegraphics[width=\linewidth]{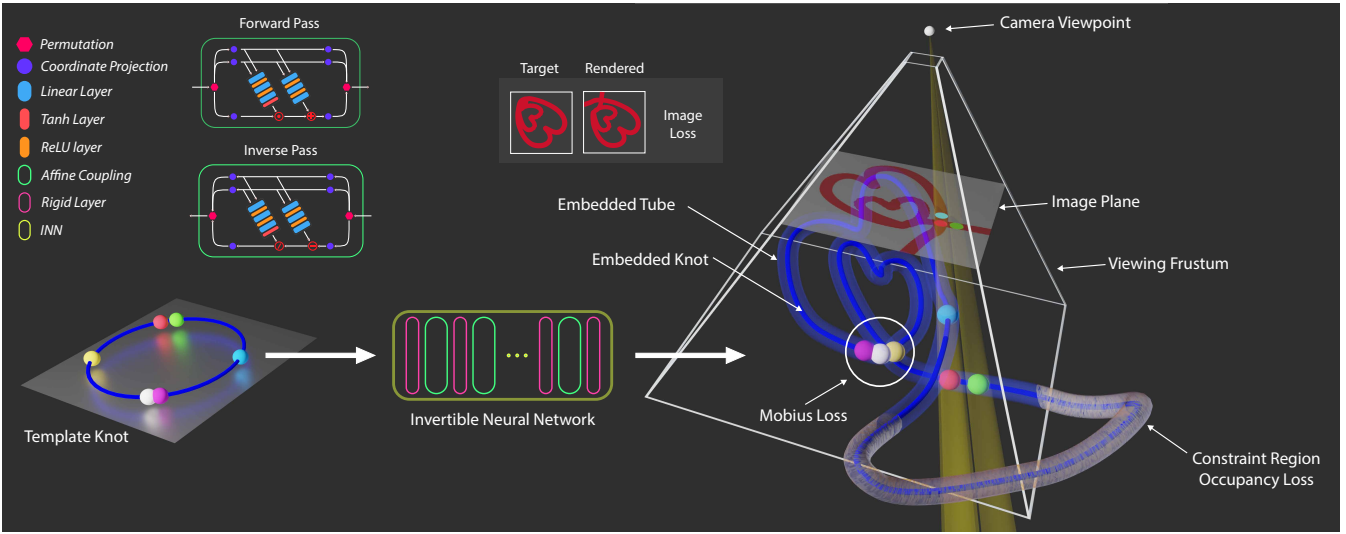}}
\caption{ Illustration showing the pipeline of our proposed framework. We forward pass a template knot through an INN consisting of an alternating sequence of affine coupling blocks and rigid transformation layers to obtain the knot representation. Given the camera viewpoint, we propose a differentiable renderer that generates the image of the knot (shown on right). Using gradient-based optimization we find the ideal knot whose rendered image is closest to the target image.}
\label{fig:pipeline}
\end{figure*}

\textbf{Inverse Rendering and Differentiable Rendering.}
Several existing inverse rendering methods recover the scene parameters from images using supervised learning \cite{bi2020deep}, \cite{yariv2020multiview} that require multi-view or single-view images either during training or inference. Recently, unsupervised and weakly supervised methods \cite{niemeyer2020differentiable}, \cite{han2020drwr}, \cite{wu2021rendering}, \cite{wimbauer2022rendering} for inverse rendering have been proposed that focus on de-rendering either general or specific object types. Differentiable Rendering algorithms depend largely on the underlying 3D data representation, either explicit ones, such as, voxels \cite{yan2016perspective}, \cite{tulsiani2018multi},  meshes \cite{loper2014opendr}, \cite{Liu_2019_ICCV} and point clouds \cite{insafutdinov2018unsupervised}, \cite{chen2021unsupervised}, \cite{yifan2019differentiable}, \cite{lin2018learning}, \cite{han2020drwr}, \cite{bi2020deep} or implicit ones \cite{liu2020dist}, \cite{niemeyer2020differentiable}, \cite{yariv2020multiview} that utilize generative models which extract the occupancy probabilities, distances, and transparencies, with respect to the surface. Our work focuses on the inverse rendering of knot-based 3D tubes from ground truth silhouette images using unsupervised learning. 

\textbf{Knot Rendering and Optimization.}
Intersection algorithms for ray tracing with curves \cite{van1985ray}, \cite{nakamaru2002ray}, \cite{reshetov2017exploiting} have been studied previously with a focus on parametric curves. J.J.Van Wijk \cite{van1985ray} formally defined and described the ray tracing intersection with the resulting shape when a sphere of changing radius is swept along a 3D trajectory. Rendering algorithms for knots have been proposed in KnotPlot \cite{scharein1998interactive}, which is a knot visualization and manipulation tool. However, these rendering methods are not differentiable. We have developed a differentiable knot renderer that facilitates search in the knot space to optimize an objective defined over the rendered image. Several existing knot energies in the literature for knot optimization, such as Möbius Energy \cite{kusner1998mobius} and Tangent-Point function \cite{yu2021repulsive}, \cite{buck1995simple} ensure the validity of the knot by penalizing self-intersections. Our approach instead has the guarantee that the represented knot is free of self-intersection, enabling optimization without the need for a penalty term.


\textbf{Template-based Shape Deformation.}
Graph-based convolution networks \cite{wang2018pixel2mesh}, MLPs \cite{tulsiani2020implicit}, \cite{groueix2018papier}, and Neural ODEs \cite{paschalidou2021neural}, \cite{gupta2020neural} have been employed before to deform a simple genus-zero shape like a sphere or an ellipsoid into a shape of arbitrary complexity. The use of Invertible Neural Networks (INNs) \cite{dinh2014nice} that utilize diffeomorphic \cite{gupta2020neural} or homeomorphic \cite{paschalidou2021neural} deformations have shown remarkable results. In our work, we utilize INNs to represent homeomorphic deformations of a template knot. Preserving the topology guarantees that there are no self-intersections in the generated knot, given that the template is self-intersection free.

\section{Proposed Approach}
\label{sec:proposedApproach}

\subsection{Problem Statement}

Let $\mathcal{C} = (t,\theta,f, z_{-}, z_{+},W,H)$ represent a pinhole camera model. Here $t=(t_x,t_y,t_z)$, $\theta=(\theta_x,\theta_y,\theta_z)$ and $f=(f_x,f_y,f_z)$ denote the camera's location, orientation, and focal lengths, respectively, with respect to the world coordinate frame. The distance of the near-clipping plane and the far-clipping plane from the camera location is denoted by $z_-$ and $z_+$, respectively. $W$ and $H$ denote the width and height of the image rendered by the camera. From here on, unless specified otherwise, we assume that all the points are defined with respect to the camera coordinate frame.

Let $\mathcal{D}=\{(x,y,z) \mid z=f_z,\ \abs{x} \leq f_x,\ \abs{y} \leq f_y\}$ denote the image plane and $\mathcal{F} = \{(x,y,z) \mid \frac{f_z}{z}(x,y,z) \in \mathcal{D},\ z_{-} \leq z \leq z_{+}\}$ denote the viewing frustum . The image pixel grid is denoted by $G=\mathcal{Z}_W \times \mathcal{Z}_H$, where $\mathbb{Z}_N=\{ z \in \mathbb{Z} \mid 0 \leq z \leq N-1 \}$, then $\mathcal{Q}=\{ (f_x\cdot(\frac{2i}{W-1}-1), f_y\cdot(\frac{2j}{H-1}-1)) \mid (i,j) \in G \}$, represents the coordinates of the pixels on the image plane.

A knot is defined as a topological embedding of the circle in $\mathbb{R}^3$.
Lets say we choose a template knot having parametric representation $\mathcal{K}: [0,1] \to \mathbb{R}^3$, with $\mathcal{K}(0)=\mathcal{K}(1)$. Let $\mathcal{H}:\mathbb{R}^3 \to \mathbb{R}^3$ be a homeomorphism. Then $\hat{\mathcal{K}} = \mathcal{H}\circ \mathcal{K}$ describes a smooth deformation of the template knot. Let $\mathcal{H}_{\phi}$ be a family of homeomorphisms, parametrized by $\phi \in \Phi$. Then, $\mathcal{K}_{\phi}=\mathcal{H}_{\phi}\circ \mathcal{K}$ represents a family of knots, parametrized by $\phi \in \Phi$. Given $0 \leq s_1 \leq s_2 \leq 1$, the arc-length of the knot segment of $\mathcal{K}_{\phi}$ between $s_1$ and $s_2$ is given by $\ell_{\mathcal{K}_{\phi}}(s_1,s_2) = \int_{s_1}^{s_2} |{\mathcal{K}^{\prime}_{\phi}}(s)| \, ds$. Let the total length of the knot be denoted as $L_{\mathcal{K}_{\phi}}$ given by $L_{\mathcal{K}_{\phi}} = \ell_{\mathcal{K}_{\phi}}(0,1)$. Let $\mathcal{T}({\mathcal{K}_{\phi}},r) = \bigcup_{s\in[0,1]}\mathcal{B}_{r}(\mathcal{K}_{\phi}(s))$ represent a tube having thickness $r$, which is obtained by sweeping a ball of radius $r$ along the knot $\mathcal{K}_{\phi}$. Here $\mathcal{B}_{r}(p)$ denotes a ball of radius $r$ centered at point $p$.

Let $\mathcal{R}: \Phi \times \mathbb{R}_{\geq 0} \times \mathcal{C}^{*} \to \mathcal{I}^{*}$ denote the rendering function, where $\mathcal{C}^{*}$ denotes the space of all pinhole cameras and $\mathcal{I}^{*} = [0,1]^\mathcal{G}$ denotes the space of all grayscale images defined on grid $\mathcal{G}$. Given $\phi \in \Phi$, $r \geq 0$ and a pinhole camera $\mathcal{C} \in \mathcal{C}^{*}$, $\hat{\mathcal{I}} = \mathcal{R}(\phi,r,\mathcal{C})$ represents the rendered image of tube $\mathcal{T}({\mathcal{K}_{\phi}},r)$ as observed by the camera $\mathcal{C}$. Given a target image $\mathcal{I} \in \mathcal{I}^{*}$, the objective is defined as 

\begin{empheq}[box=\fbox]{align}
\min_{\phi \in \Phi} \quad {||\mathcal{I} - \mathcal{R}(\phi,r,\mathcal{C})||}_2^2  \quad \text{s.t.} \label{eq:objectivefn}\\ 
\Lambda(\mathcal{T}({\mathcal{K}_{\phi}},r))=\emptyset,  & \quad B_{\mathcal{K}_{\phi}} \leq B_{0}, \nonumber \\
\mathcal{T}({\mathcal{K}_{\phi}},r) \subseteq \Omega_{0}, & \quad L_{\mathcal{K}_{\phi}} \leq L_{0} \nonumber
\end{empheq}

Here, $\Lambda(\mathcal{T}({\mathcal{K}_{\phi}},r))$ denotes the self-intersections in the tube, $B_{\mathcal{K}_{\phi}}$ denotes the bending (curvature) in the knot and $L_{\mathcal{K}_{\phi}}$ denotes the total length of the knot. $\Omega_{0}$ represents the constraint region in space inside which the tube is supposed to lie, $B_{0}$ is the maximum allowed bending, and $L_{0}$ is the maximum allowed length of the knot.

\subsection{Parametric Family of Knots}

$\mathcal{H}:\mathbb{R}^3 \to \mathbb{R}^3$ is said to be a homeomorphism, if $\mathcal{H}$ is bijective and bicontinuous. We use Invertible Neural Networks (INN) to represent $\mathcal{H}_{\phi}$ , where $\phi$ denotes the INN parameters. We design an INN consisting of an alternating sequence of Affine Coupling Layers \cite{dinh2016density} and Rigid transformation layers. The INN is invertible by design, guaranteeing  bijectivity of $\mathcal{H}_{\phi}$ and the INN being a compositional function of continuous layers guarantees the bicontinuity of $\mathcal{H}_{\phi}$.

Using an INN-based parametric family of homeomorphisms to search in the space of knots has several advantages. The INN being differentiable facilitates efficient gradient-based optimization for searching the optimal knot. By choosing a template knot $\mathcal{K}$, the use of homeomorphism guarantees that the search space is constrained to knots having the same knot type as that of $\mathcal{K}$. Thus the desired knot type can be fixed by appropriately choosing the template knot. 
An embedded knot by definition is free of self-intersection. If it is ensured that the template used is a valid knot, then it is guaranteed that $\mathcal{K}_{\phi}$ is a valid knot for any choice of $\phi \in \Phi$. Thus every $\mathcal{K}_{\phi}$ is free of self-intersection since a homeomorphism preserves the topological properties. Thus our approach does not require any additional projection step or the use of loss functions to ensure the validity of the knots.

\begin{figure}
\centering
\centerline{\includegraphics[width=\linewidth]{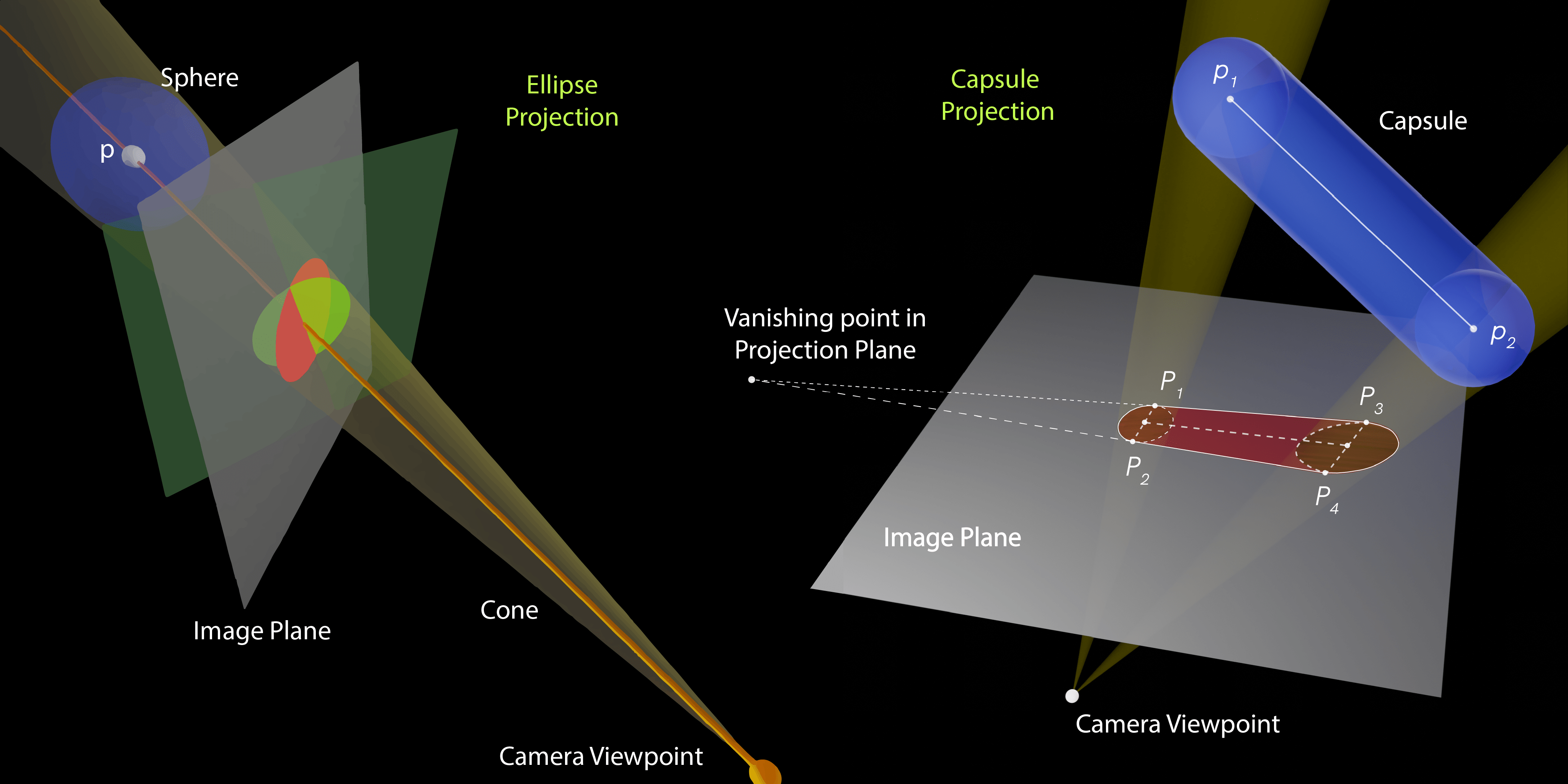}}
\caption{Illustration of the Ellipse Renderer ($\mathcal{R_E})$) and the Capsule Renderer ($\mathcal{R_C})$).}
\label{fig:CapsuleProjection}
\end{figure}

Our approach works for any choice of template knot. It is general in the sense that, if we choose open curves instead of closed knots, then we can search in the space of open curves and render open tube structures. However, here we restrict our focus only to knots, specifically the trivial knot, whose template is represented as $\mathcal{K}:[0,1] \to \mathbb{R}^3$, $\mathcal{K}(s)=(\cos s,\sin s, 0)$. We demonstrate that even a trivial knot is powerful enough to represent complex images. In Fig. \ref{fig:pipeline}, the INN and the affine coupling  blocks are illustrated.

\subsection{Differentiable Silhouette Rendering of Knots}

We propose two algorithms for the rendering function $\mathcal{R}(\phi,r,\mathcal{C})$, both of which being differentiable, facilitates the end-to-end optimization of the objective defined in Eq. \ref{eq:objectivefn}. 

\noindent \textbf{Ellipse Renderer ($\mathcal{R_E}$).} We first randomly sample ${N}$ points on the template unknot $\mathcal{K}$ to obtain the set $\mathcal{P_\mathcal{K}}$. We forward pass the template points through the INN $\mathcal{H}_{\phi}$ to obtain points on $\mathcal{K}_{\phi}$, denoted as the set  $\mathcal{P}_{\mathcal{K}_{\phi}} = {\mathcal{H}_{\phi}(\mathcal{P_\mathcal{K}})}$. 
The sphere of radius $r$ centered at a point $p = (l, m, n) \in \mathcal{P}$ will project an ellipse, $\mathcal{E}_p$, on the image plane.
Let $\mathcal{N}_p$ be the cone induced by $p$ such that its apex is at the origin, height is $\norm{p}_2$ and radius is $r$. 
The projected ellipse $\mathcal{E}_p$ is the intersection of the cone $\mathcal{N}_p$ with the image plane. 
The equation of $\mathcal{E}_p$ is given by $Ax^2 + Bxy + Cy^2 + Dx + Ey + F = 0$, where $A = l^2 - k^2$, $B = 2lm$, $C = m^2 - k^2$, $D = 2f_zln$, $E = 2f_zmn$, $F = f_z^2(n^2 - k^2)$, and $ k^2 = \frac{(l^2 + m^2 + n^2)^2}{l^2 + m^2 + n^2 + r^2}$.
Projecting ellipses about each point in $\mathcal{P_K}$ on the image plane gives a good approximation of the $\mathcal{T}({\mathcal{K}_{\phi}},r)$ when $N$ is sufficiently large. However, if $N$ is very small, or if the distance between consecutive sample points is large, then it might result in an undesirable discontinuity in the rendered image. To avoid this, we propose another renderer which has additional computational cost but is more accurate.  

\begin{figure*}[!t]
\centering
    \begin{minipage}[b]{0.49\linewidth}
        \centerline{\includegraphics[width=\linewidth]{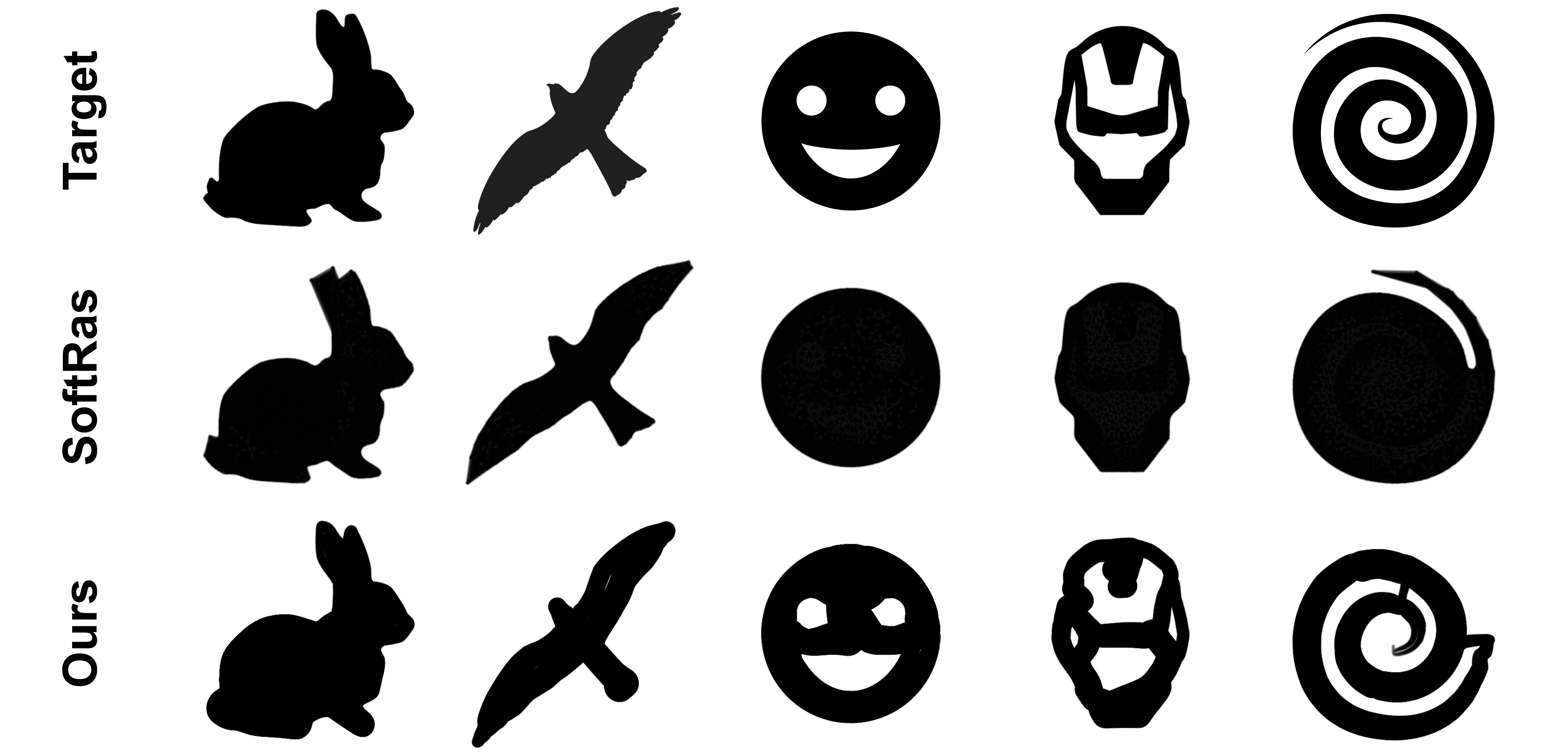}}
        \centerline{(a) With Constraints}\medskip
   \end{minipage}
   \begin{minipage}[b]{0.49\linewidth}
        \centering
        \centerline{\includegraphics[width=\linewidth]{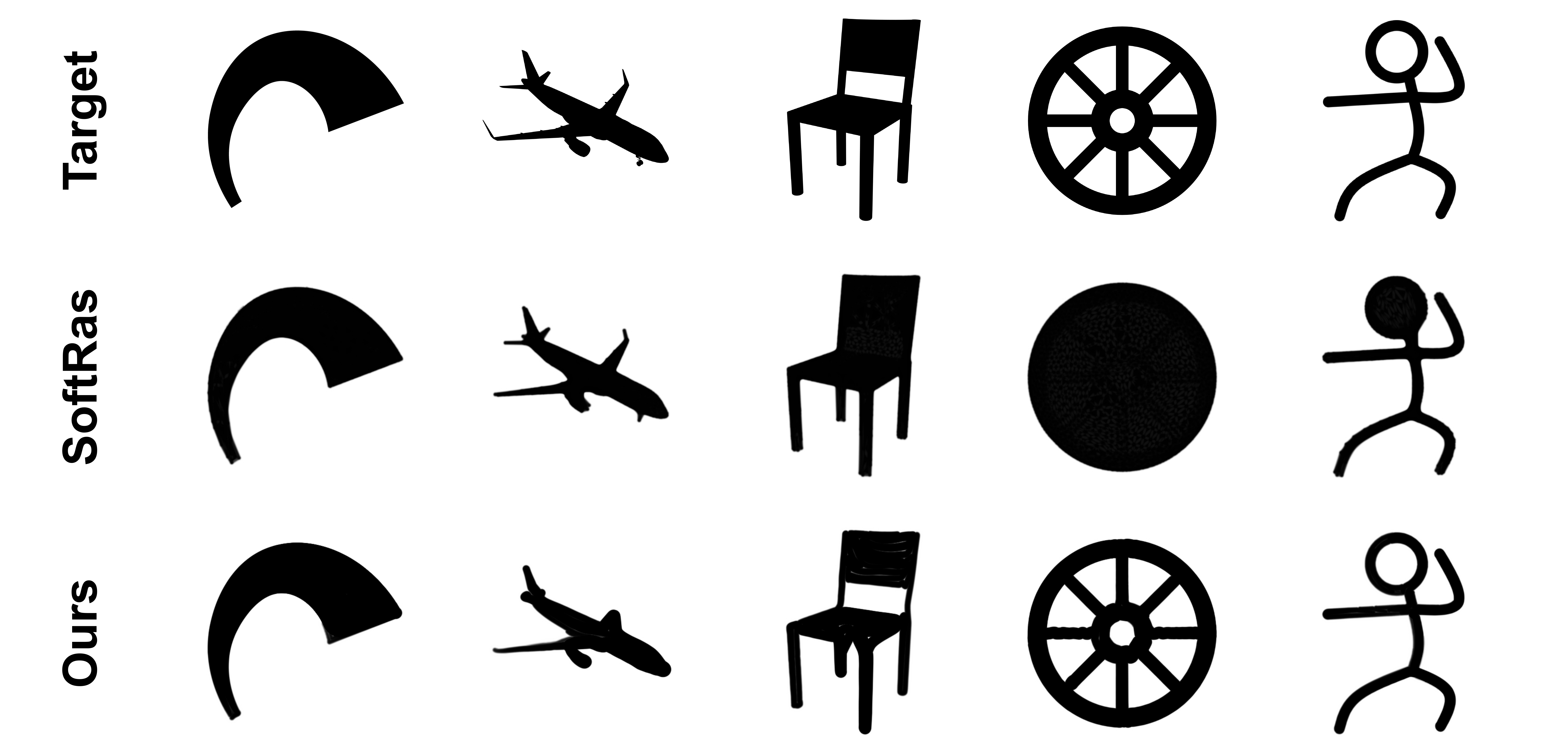}}
        \centerline{(b) Without Constraints}\medskip
   \end{minipage}
    \captionof{figure}{Rendering of the 3D tube embeddings obtained using our method (bottom row) and the rendering of the sphere mesh deformation using SoftRas \cite{Liu_2019_ICCV} (middle row) for the corresponding ground truth silhouettes (top row). For results in (a) the physical realizability constraints are applied, whereas in (b) these constraints are relaxed.}
    \label{fig:SVSK}
\end{figure*}


\noindent \textbf{Capsule Renderer ($\mathcal{R_C})$.} We consider a cylindrical capsule between consecutive points on the knot $\mathcal{K}_\phi$, having spherical ends formed by spheres of radius $r$ centered at those points. Let the capsule have a projection $\mathcal{U}_p$ on the image plane.  $\mathcal{U}_p$ is the union of the projections of the two spheres at the ends of the capsule and the projection of the central cylinder plane joining them. The analytical equations and derivations of the projected ellipse and projected capsule have been included in the supplementary material. 
Let ${S}_{\mathcal{U}_p}: {\mathbb{R}}^2 \to \mathbb{R}$ be the signed distance function of the projected capsule $\mathcal{U}_p$. The occupancy function of $\mathcal{U}_p$ on the image plane can then be defined as $\mathcal{O}_{\mathcal{U}_p}(q) = \sigma(\tau \cdot {S_{\mathcal{U}_p}}(q))$, where $\sigma$ is the sigmoid function. The parameter $\tau$ is the hardness factor which controls how rapidly the occupancy function changes near the boundary. Let us consider a pixel $q \in \mathcal{Q}$ having grid coordinates $(i,j)$. The pixel value of $q$, denoted as $\hat{\mathcal{I}}(i,j)$ can be computed as the maximum of the occupancy values across all the projected ellipses, given by, 
$\mathcal{\hat{\mathcal{I}}}(i,j) = \max_{p \in \mathcal{P_F}} \mathcal{O}_{\mathcal{U}_p}(q)$. The differentiable rendering process is illustrated in Fig. \ref{fig:CapsuleProjection}.

\subsection{Embedding Search of Knots}

Given the objective defined in Eq.\ref{eq:objectivefn}, the rendering function and the representation of knots being differentiable enable us to use gradient descent-based methods for optimization. The objective and the constraints are represented in terms of loss functions which are minimized. First, we define some preliminary concepts needed to describe the loss functions.
Let $p_1=\mathcal{K}_{\phi}(s_1)$ and $p_2=\mathcal{K}_{\phi}(s_2)$ denote two points on the knot, where $0\leq s_1 \leq s_2 \leq 1$. Let $d(p_1,p_2)=\norm{p_1 - p_2}_2$ denote the Euclidean distance between the two points.
Let $g(p_1, p_2) = \min(\ell_{\mathcal{K}_{\phi}}(s_1, s_2),\; L_{\mathcal{K}_{\phi}} - \ell_{\mathcal{K}_{\phi}}(s_1, s_2))$ denote the geodesic distance between the two points.
The loss functions are illustrated in Fig. \ref{fig:pipeline}.\\

\textbf{Image Loss ($\mathcal{L}_I$)}:
This loss is used to make the rendered image $\hat{\mathcal{I}}$ similar to the target image $\mathcal{I}$ and is given by
\begin{equation}
\begin{aligned}
    \mathcal{L}_{I} = \norm{\mathcal{I} - \hat{\mathcal{I}}}_2^2
\end{aligned}
\end{equation}

\textbf{Length Loss ($\mathcal{L}_L$)}: 
As the radius of the knot is constant throughout, the material cost of the tube is proportional to the length of the knot. The material cost budget corresponds to a maximum allowed length $L_{0}$. The length loss adds a penalty whenever knot length exceeds $L_{0}$, given by
 \begin{equation}
     \begin{aligned}
          \mathcal{L}_{L} = \max( L_{\mathcal{K}_{\phi}} - L_{0}, 0)
     \end{aligned}
 \end{equation}
The set $\mathcal{P_K}$ is obtained by sampling a random ring-graph from the circle, which is then forward passed through the INN $\mathcal{H}_{\phi}$ to obtain $\mathcal{P}_{\mathcal{K}_{\phi}}$ . The length of edges in
the deformed ring are added up to obtain the knot length $L_{\mathcal{K}_{\phi}}$.

\textbf{M\"{o}bius Loss ($\mathcal{L}_M$)}:
Even though the knot $\mathcal{K}_{\phi}$ is guaranteed to have no self-intersection in our representation, the tube $\mathcal{T}({\mathcal{K}_{\phi}},r)$ can have self-intersections. In order to penalize tube self-intersections, we define a loss based on the M\"{o}bius Energy. The M\"{o}bius Energy between two knot points is defined as $\mathcal{M}(u,v) = \frac{1}{d(u,v)^2} - \frac{1}{g(u,v)^2}$. 
We define M\"{o}bius loss on a randomly sampled set $\mathcal{B}_M = \{ (u,v) \mid u,v \in \mathcal{P_{K_\phi}},\, u \neq v \}$ as
\begin{equation}
\begin{aligned}
    \mathcal{L}_{M} = \frac{1}{|\mathcal{B}_M|}\left( \sum_{(u,v)\in \mathcal{B}_M} \mathcal{M}(u,v) \cdot \max(2r - d(u,v), 0) \right)
\end{aligned}
\end{equation}
This loss penalizes those pairs of points for which the geodesic distance is much larger than the euclidean distance and the euclidean distance is less than twice the radius (implying tube self-intersection).

\textbf{Occupancy Loss ($\mathcal{L}_R$)}:
An artist or an architect while realizing such a tube structure might have space constraints and would like the structure to be bounded within a predefined constraint region $\Omega_0$.
Let $S_{\Omega_0}: \mathbb{R}^3 \to \mathbb{R}$ be the signed distance function of $\Omega_0$, which is positive for the inside points and negative for the outside points.
The occupancy Loss penalizes the tube points that go outside the constraint region $\Omega_0$ and is defined as
\begin{equation}
\begin{aligned}
    \mathcal{L}_{R} = \sum_{p \in \mathcal{P}_{\mathcal{K_\phi}}} \max(r - S_{\Omega_0}(p), 0)
\end{aligned}
\end{equation}
This loss adds penalty whenever a sphere of radius $r$ about a knot point does not completely lie within $\Omega_0$.

\textbf{Bending Loss ($\mathcal{L}_R$)}:
Based on the material to be used for the tube, there might be restrictions on the extent to which the tube can physically bend. Let  $B_0$ be the maximum bending that a tube can physically attain at any point. The bending loss penalizes those points on the knot whose squared curvature, given by $B_{\mathcal{K}_{\phi}} (p)$ exceeds the maximum allowed bending $B_0$ and is defined as
\begin{equation}
\begin{aligned}
    \mathcal{L}_{B} = \sum_{p \in \mathcal{P}_{\mathcal{K_\phi}}} \max(B_{\mathcal{K}_{\phi}}(p) - B_{0}, 0)
\end{aligned}
\end{equation}



The total loss function is a weighted sum given as $\mathcal{L} = w_{img}\cdot\mathcal{L}_{I} + w_{len}\cdot\mathcal{L}_{L} + w_{mob}\cdot\mathcal{L}_{M} + w_{occ}\cdot\mathcal{L}_{R} + w_{bend}\cdot\mathcal{L}_{B}$. The hyperparameters used in our framework, the computing infrastructure used to run the experiments and the details of the INN architecture are included in the supplementary material.
\section{Results and Discussion}
\label{sec:resultsAndDiscussion}

We create a test bed of target silhouette images and demonstrate in Sec. \ref{subsec:qualitativeresults} that our method obtains impressive results. 
We also conduct experiments (Sec. \ref{subsec:experiments} and \ref{subsec:ablationrenderer}) which indicate the effectiveness of the proposed loss functions and the renderer. A real world demonstration is shown in Sec. \ref{subsec:realworlddemo}.

\subsection{Inverse perceptual art results}
\label{subsec:qualitativeresults}

The proposed framework was tested on silhouette images having varying levels of complexity, under different settings.

\noindent \textbf{Single Viewer, Single Knot.}
In this problem setting, there is a single viewer configuration with a single target image. The objective is to find the embedding of a single knot whose projected image resembles the target image as illustrated in Fig. \ref{fig:ironman_viviani}(a). 
To the best of our knowledge, no existing works in the literature address the problem of knot-based inverse perceptual art (mentioned in Eq. \ref{eq:objectivefn}). This prohibits us from performing exhaustive comparative analysis with other methods. 
However, the work of SoftRas \cite{Liu_2019_ICCV}, which instead of tube embedding, uses sphere mesh deformation, closely resembles our work, with whom we compare our results, as shown in Fig. \ref{fig:SVSK}. 
In SoftRas, due to deformation of a sphere mesh, there is difficulty in forming image shapes with several holes. In contrast our method is able to generate complex shapes with complex topologies.

\noindent \textbf{Single Viewer, Multiple Knots.}
In this problem setting, there is a single viewing location and multiple knots. The objective is to find the embedding of all the knots so that the projected image resembles the target image. Fig. \ref{fig:Teaser}(a) shows the result of a single light source casting shadows from multiple knots. And when a viewer is placed at the location, all the knots together form a single image.

\begin{figure}
\centering
\centerline{\includegraphics[width=\linewidth]{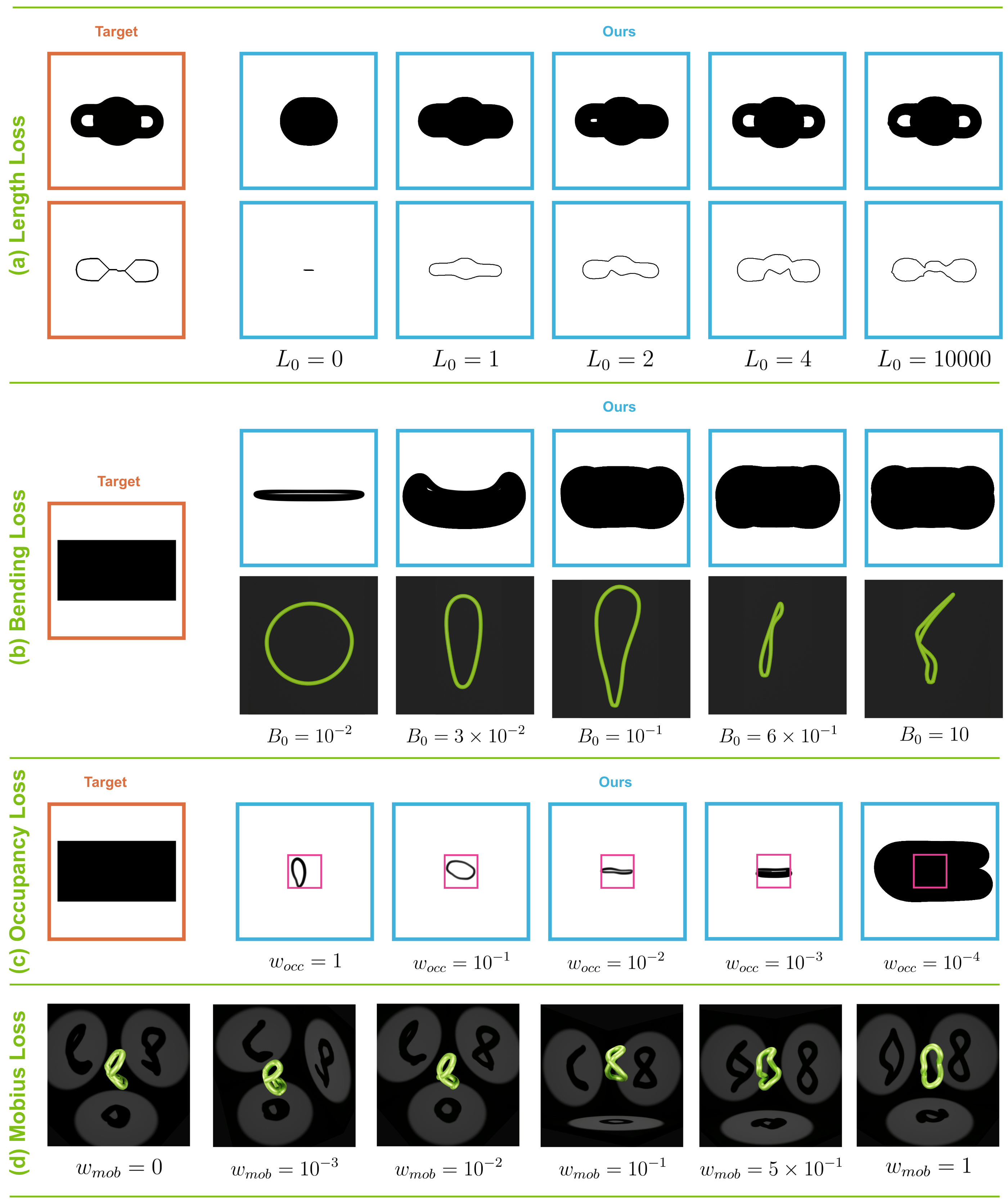}}
\caption{Ablation experiments of proposed loss functions.}
\label{fig:ablation}
\end{figure}

\noindent \textbf{Multiple Viewer, Single Knots.}
In this case, there are multiple viewing configurations and a target image corresponding to each view. The embedding of a single knot needs to be searched, whose projected image from each view matches the respective target image.
As demonstrated in Fig. \ref{fig:ironman_viviani}(b) and Fig. \ref{fig:ablation}(d), there are three target silhouette images representing projections on three perpendicular image planes. These three images are the orthographic projections of Viviani's curve. Viviani's curve has a self-intersection in the middle and is thus not a knot. Our method instead learns a knot embedding that is physically realizable, free of self-intersection, whose perspective projections on the image plane resemble the orthographic projections of Viviani's curve. 

\noindent \textbf{Single Viewer (single target video), Spatio-temporal Knots.}
In this scenario, given a temporal sequence of target images, the objective is to find the temporal sequence of embeddings of a single knot, so that at a given time instant, the rendering of the knot embedding is similar to the target image at the corresponding time instant. In this case, additional regularization is added to ensure smooth spatio-temporal deformation of the knot. In Fig. \ref{fig:Teaser}(b), all the knots are constrained to be smooth temporal deformations of each other. The animation depicting this temporal deformation of the knots creates the perception of a person dancing. Thus, in this case, the perceptual art also has a temporal aspect. The animation is available in the supplementary material. 

Our approach learns knot embeddings whose projected images closely resemble the target silhouette images in all the scenarios, proving the effectiveness of our approach. The knot-based tube embeddings when rendered in 3D, were also found to satisfy the physical constraints specified.  
More results along with the time evolution of the 3D knot embeddings during the optimization process are available in the supplementary material.

\subsection{Ablation Study of Loss Functions}
\label{subsec:experiments}

We conduct experiments to investigate the importance of each loss function proposed in our approach. Since image loss is the primary objective, keeping it weight fixed at $w_{img} = 1$, we vary the other loss function's hyperparameters and observe the effects.

\noindent \textbf{Length Loss.} Fig. \ref{fig:ablation}(a) shows the rendered image and its skeleton corresponding to a given ground truth, for different considered values of the maximum allowed length, $L_0$. For smaller values of $L_0$, the proposed algorithm tries to render an image as close to the ground truth, while respecting the budget constraint set by $L_0$. When large values of $L_0$ are considered, 3D knot embedding is given more freedom to deform in space and successfully generates the desired image.


\noindent \textbf{Bending Loss.} In Fig. \ref{fig:ablation}(b), given the target image of a rectangle, the maximum allowed bending $B_0$ is varied. For low values of $B_0$, the bending loss forces the knot to have the least amount of curvature, resulting in the symmetric circular shape. Only when higher value of bending is allowed, the knot starts to deform, and eventually for a sufficiently large value of $B_0$, deforms to result in an image that closely resembles a rectangle.

\noindent \textbf{Occupancy Loss.} In Fig. \ref{fig:ablation}(c), the target image is a rectangle and the pink sub-region within the image indicates the constraint region. When the occupancy loss weight $w_{occ}$ is higher, the knot is forced to stay confined with the constrained region. Only when $w_{occ}$ is made sufficiently small, the other losses start to dominate and defy the region constraints. This indicated that the occupancy loss successfully forces the knot to stay within a user defined region.

\noindent \textbf{M\"{o}bius Loss.} The effect of M\"{o}bius loss is investigated in Fig. \ref{fig:ablation}(d), in the multi-view, multi-target setting. The target images are chosen in such a way that the optimal solution necessitates a self-intersection in the tube. For smaller values of $w_{mob}$, we see that the M\"{o}bius loss is not given significance and the resulting configuration of tube indeed has self-intersections. Only when the weight $w_{mob}$ is increased sufficiently, does the M\"{o}bius loss component become strong enough to avoid self-intersections. This demonstrates that M\"{o}bius loss indeed avoids self-intersection in the tube.

\begin{figure}[!t]
\centering
\centerline{\includegraphics[width=\linewidth]{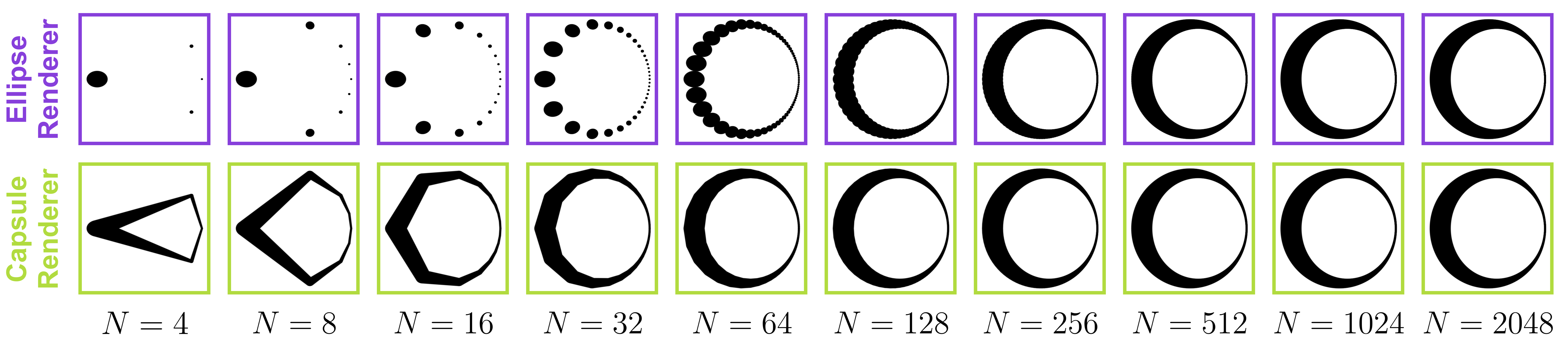}}
\caption{Renderings for varying number of $N$.}
\label{fig:renderer_vary_N}
\end{figure}

\begin{figure}[!t]
\centering
\centerline{\includegraphics[width=\linewidth]{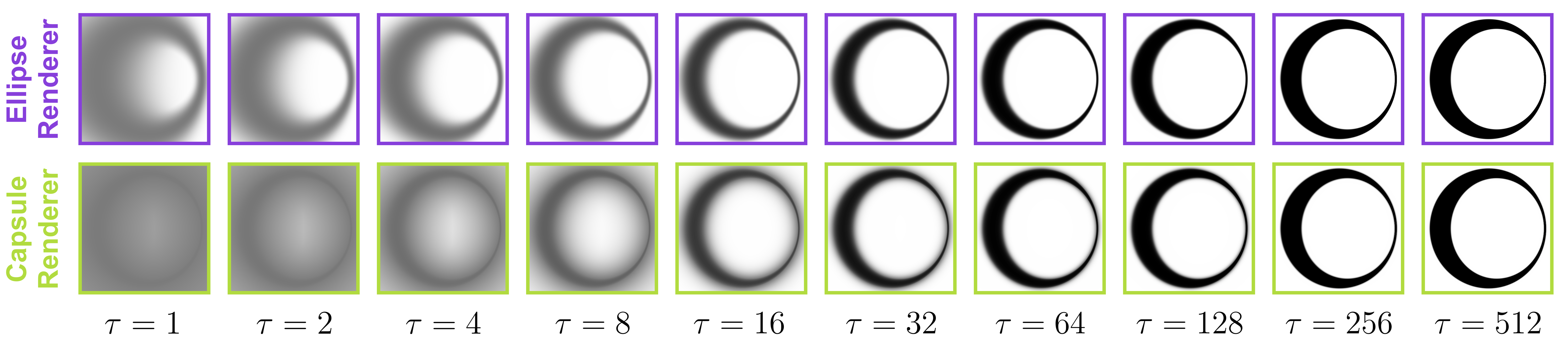}}
\caption{Renderings for varying value of $\tau$.}
\label{fig:renderer_vary_tau}
\end{figure}

\subsection{Differentiable Renderer Experiments}
\label{subsec:ablationrenderer}

We have proposed two methods for rendering the tube: Ellipse Renderer ($\mathcal{R_E}$) and Capsule Renderer ($\mathcal{R_C}$). In this section we discuss the scenario when one would be preferred over the other.
In Fig. \ref{fig:renderer_vary_N}, for a fixed circle in 3D, we vary the number of sample points $N$ and see its effect on $\mathcal{R_E}$ and $\mathcal{R_C}$. When $N$ is large, both produce the desired results. However, when $N$ is small, such as $N=64$, $\mathcal{R_E}$ generates a poor approximation, whereas $\mathcal{R_C}$ produces a much better approximation of the actual image. This demonstrates that $\mathcal{R_C}$ is superior than $\mathcal{R_E}$, when $N$ is small. Thus $\mathcal{R_C}$ is particularly advantageous, if the INN $\mathcal{H}_{\phi}$ has many layers, in which case querying points from the knot (forward pass) is computationally expensive. If the forward pass through INN $\mathcal{H}_{\phi}$ is computationally cheap, then querying points from the knot is inexpensive, making the computations involved in capsule rendering the main bottleneck. In such a case, $\mathcal{R_E}$ would be preferred over $\mathcal{R_C}$.

In Fig. \ref{fig:renderer_vary_tau}, for both $\mathcal{R_E}$ and $\mathcal{R_C}$, we vary the hardness factor $\tau$. Making $\tau$ large generates a rendering close to the desired image. But large $\tau$ also results in exploding gradients making the optimization process numerically unstable. For $\tau$ extremely small, the optimization is stable, but the rendered image is less sharp, producing knot configuration whose actual image is different from the desired target. In our experiments we use the value of $\tau=100$, and observe that it gives good result, while also being numerically stable.

\subsection{Real World Demonstration}
\label{subsec:realworlddemo}

\begin{figure}
\centering
\centerline{\includegraphics[width=\linewidth]{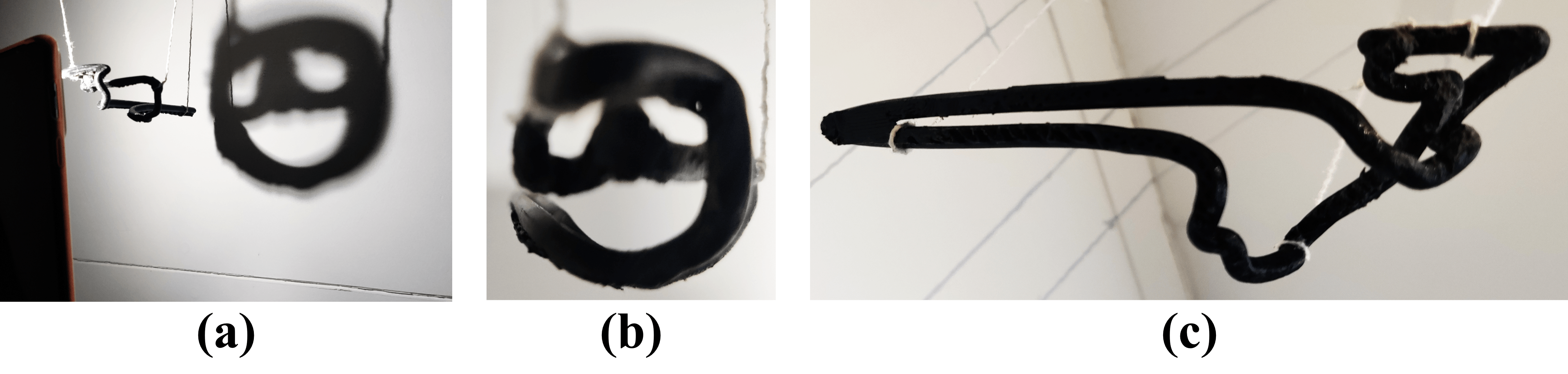}}
\caption{Real-world demonstration of our approach.}
\label{fig:realworld}
\end{figure}

Given the target image of a smiley (third column in Fig. \ref{fig:SVSK}), our approach generates a tube structure, which we fabricate using 3D printer, shown in Fig. \ref{fig:realworld}(c). We cast light on this structure (Fig. \ref{fig:realworld}(a)) whose shadow resembles a smiley and when viewed from the specified location (Fig. \ref{fig:realworld}(b)) it appears like a smiley.
\section{Conclusion and Future Work}
\label{sec:conclusion}

In this work, we have proposed an end-to-end differentiable optimization framework for searching 3D knot-based tube embedding that, when projected onto the image plane of a user-specified camera, forms a silhouette image similar to the input target image. In order to represent the knot embeddings, we have used invertible neural networks (INNs) as a parametrized family of homeomorphisms and created a differentiable silhouette renderer that renders the tube. We also present several loss functions to ensure that the generated tubular structure is physically realizable. We present the power of a simple unknot being used as the template knot and how its deformation through homeomorphism can be derendered even in the case of complex images. The inverse knot rendering from single-view and multiple-view sets of silhouette images is represented for any given camera attribute. In future work, we aim to experiment with other formulations of inverse perceptual art. For instance, in the current setting, the viewer configuration and radius is fixed for the desired target image. A variant in which the viewer configurations are left unspecified, would be an interesting problem. In such a variant, the proposed approach will also find the viewer configuration along with the knot embedding and the thickness of the tube. The current differentiable renderer is silhouette based. In future, we propose to develop realistic shading-based renderers for knots, which can help in the derendering of real images of knots.

\bibliography{aaai24}

\newpage

\section{Capsule projections}

The cylindrical capsule between consecutive points, $p_1 = (l_1, m_1, n_1) \in \mathcal{P}_{\mathcal{K}_{\phi}}$ and $p_2 = (l_2, m_2, n_2) \in \mathcal{P}_{\mathcal{K}_{\phi}}$
on the knot $\mathcal{K}_\phi$, having spherical ends formed by spheres of radius $r$ centered at those points, has a projection $\mathcal{U}_{p_{1,2}}$ on the projection plane, $z = f_z$. $\,\,\mathcal{U}_{p_{1,2}}$ is the union of the projections of the two spheres at the ends of the 3D capsule and the projection of the central cylinder plane joining them. We first derive the equation for the projection of the spherical ends.

\subsection{Equation of Projection of Spheres}

The sphere of radius $r$ centered at a point $p = (l, m, n) \in \mathcal{P}_{\mathcal{K}_{\phi}}$ will project an ellipse on the image plane. Let $\mathcal{N}_p$ be the cone induced by $p$ such that its apex is at the origin, height is $\norm{p}_2$ and radius is $r$. The projected ellipse $\mathcal{E}_p$ is the intersection of the cone $\mathcal{N}_p$ with the image plane. We will first derive the equation of the cone $\mathcal{N}_p$.

Let us consider a general point on $\mathcal{N}_p$, given by $n = (x, y, z)$. Let the aperture of the cone be $\text{\textit{2}}\theta$. The cosine of the angle $\theta$ can be computed by taking the dot product of the vectors representing the cone's axis, given by $({l}, {m}, {n})$, and the line joining the cone's apex to the considered point $n$, given by $({x}, {y}, {z})$, such that

\begin{equation}
\label{cos1}
    cos(\theta) = \frac{lx+my+nz}{(\sqrt{l^2 + m^2 + n^2})(\sqrt{x^2 + y^2 + z^2})}
\end{equation}

Since the considered cone is right circular, the cosine of the angle $\theta$ can also be computed as, 
\begin{equation}
\label{cos2}
cos(\theta) = \frac{\sqrt{l^2 + m^2 + n^2}}{\sqrt{l^2 + m^2 + n^2 + r^2}}
\end{equation}

The equation of $\mathcal{N}_p$ obtained by solving the above two equations, Eq. \ref{cos1} and Eq. \ref{cos2} simultaneously is as follows, 
\begin{multline}
    (l^2 - k^2) \; x^2 \;+\; (m^2 - k^2) \; y^2 \;+\; (n^2 - k^2) \;z^2 \\ \;+\;  2lmxy \;+\; 2mnyz \;+\; 2lnxz \; = \; 0 \\ where \;\; k^2 = \frac{(l^2 + m^2 + n^2)^2}{l^2 + m^2 + n^2 + r^2}
\end{multline}

The equation of the ellipse, $\mathcal{E}_p$, can then be obtained by simultaneously solving the projected plane $z = f_z$ and $\mathcal{N}_p$, given by
\begin{multline}
\label{ellipse}
    (l^2 - k^2) \; x^2 \;+\; 2lmxy \;+\; (m^2 - k^2) \; y^2 \;+\; \\ 2lnxf_z \;+\; 2mnyf_z \;+\; (n^2 - k^2) \;f_z^2 \; = \; 0 \\ \forall (x, y) \;\; s.t. \;\; \abs{x} \leq f_x \; , \;\; \ \abs{y} \leq f_y
\end{multline}

Eq. \ref{ellipse} is in the form of the general equation of an ellipse given by $Ax^2 + Bxy + Cy^2 + Dx + Ey + F = 0$
, where $A = l^2 - k^2$, $B = 2lm$, $C = m^2 - k^2$, $D = 2f_zln$, $E = 2f_zmn$, $F = f_z^2(n^2 - k^2)$.

\begin{figure}[!t]
    \centerline{\includegraphics[width=1.0\linewidth]{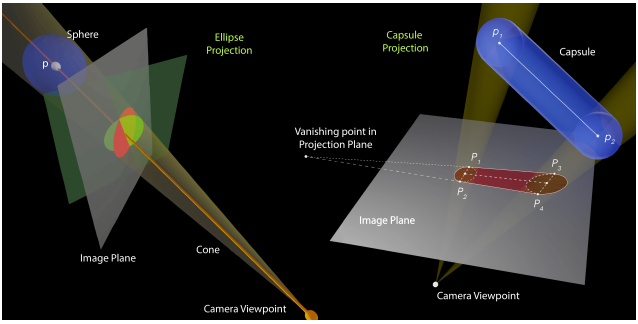}}
    \captionof{figure}{Illustration of the Ellipse Renderer ($\mathcal{R_E})$ and the Capsule Renderer ($\mathcal{R_C}$).}
    \label{fig:CapsuleProjection2}
\end{figure}%

\subsection{Equation of Projection of Cylinder}

 We now derive the equation for the projection of the central cylinder connecting the spherical ends of the considered capsule, $\mathcal{U}_{p_{1,2}}$. From Fig. \ref{fig:CapsuleProjection2}, it can be observed that the 2D projection of the cylinder will be a quadrilateral $P_1P_2P_3P_4$ such that the one pair of its opposite sides, $P_1P_3$ and $P_2P_4$, is tangent to both the projected ellipses $\mathcal{E}_{p_1}$ and $\mathcal{E}_{p_2}$, while the other pair of sides, $P_1P_2$ and $P_3P_4$, lie completely inside each of the two ellipses. The tangent lines corresponding to the first pair of sides can also be observed to intersect at a vanishing point $v = (x_0, y_0)$ on the projection plane. Given the equation of the projected ellipses $\mathcal{E}_{p_1}$ and $\mathcal{E}_{p_2}$, and the point $v$, we find the points of contact of the tangents on both the ellipses and hence obtain the four corner points of the projected quadrilateral.

The vanishing point $v = (x_0, y_0)$ on the image plane can be viewed as the projection of a point on the line ${L}(\lambda) = {p_1} + \lambda {D}$ parameterized by a constant $\lambda$, where ${D}$ is the direction vector joining $p_1$ and $p_2$, for which
\begin{equation}
    v = \underset{\lambda \rightarrow \pm \infty} {lim} \frac{f_z {L}(\lambda)}{Z}
\end{equation}
\begin{equation}
    \Rightarrow x_0 = f_z \left( \frac{l_1 - l_2}{n_1 - n_2} \right) \text{ ,  } y_0 = f_z \left( \frac{m_1 - m_2}{n_1 - n_2} \right)
\end{equation}

Consider the equation of the projected ellipse given in Eq. \ref{ellipse} in its general form $\mathcal{E}_p \equiv Ax^2 + Bxy + Cy^2 + Dx + Ey + F = 0$. Let the tangents to $\mathcal{E}_{p_1}$ from $v$ be of the form $y = mx + c$, where m is the slope and c is the intercept of the tangent in the coordinate space of the projection plane. Since $v$ is also a point on the tangent, the tangent lines become 
\begin{equation}
\label{tangent}
    y = m(x-x_0) + y_0
\end{equation}
Solving the equation of the tangent line and the equation of $\mathcal{E}_{p_1}$ we obtain the following quadratic equation in variable $x$,
\begin{multline}
\label{quad1}
    (A + Bm + Cm^2) \; x^2 + \\ (-Bmx_0 + By_0 - 2Cm^2x_0 + 2Cmy_0 + Em + D) \; x + \\ (Cm^2x_0^2 + Cy_0^2 - 2Cmx_0y_0 - Emx_0 + Ey_0 + F) = 0
\end{multline}
For each of the two values of $m$ corresponding to the tangent lines, the above equation yields two points on the ellipse. However, since the lines are tangents touching the ellipse at only one point, both these points should be the same. Thus, the determinant of the above equation will be zero, giving us the following quadratic in $m$:

\begin{multline}
\label{quad2}
    \alpha \, m^2 + \beta \, m + \gamma = 0, \\ 
    \alpha = (Bx_0 - 2Cy_0 - E)^2 - 4Cx_0(By_0 + D) \\ - 4C(Cy_0^2 + Ey_0 + F) -4ACx_0^2  \\+ 4B(2Cx_0y_0 + Ex_0)
    \\ \beta = -2(Bx_0 - 2Cy_0 - E)(By_0 + D) \\-  4B(Cy_0^2 + Ey_0 +F) \\+ 4A(2Cx_0y_0 + Ex_0)
    \\ \gamma = (By_0 + D)^2 - 4A(Cy_0^2 + Ey_0 + F) \, \phantom{spac}
\end{multline}

The roots of the above quadratic can be found using the quadratic formula that gives $m = \frac{-\beta \pm \sqrt{\beta^2 - 4\alpha\gamma}}{2\alpha}$. Let the two values of $m$ be $m_1$ and $m_2$. Using Eq. \ref{tangent} and Eq. \ref{quad1}, we can find the two points of contact on $\mathcal{E}_{p_1}$, given by $P_1$ and $P_2$. Similarly, by deriving Eq. \ref{quad1} for $\mathcal{E}_{p_2}$, we can find the other set of points, $P_3$ and $P_4$.

The coordinates of the four points, $P_1, P_2, P_3, P_4$ can be used can be used to find the equations of the lines forming the sides of the quadrilateral. The equations of the ellipse and the quadrilateral allow us to find the signed distance function of the projection of the capsule, ${S}_{\mathcal{U}_{p_{1,2}}}$, and hence the occupancy function $\mathcal{O}_{\mathcal{U}_{p_{1,2}}}(q)$.

\begin{figure}[!t]
\centering
\centerline{\includegraphics[width=1.0\linewidth]{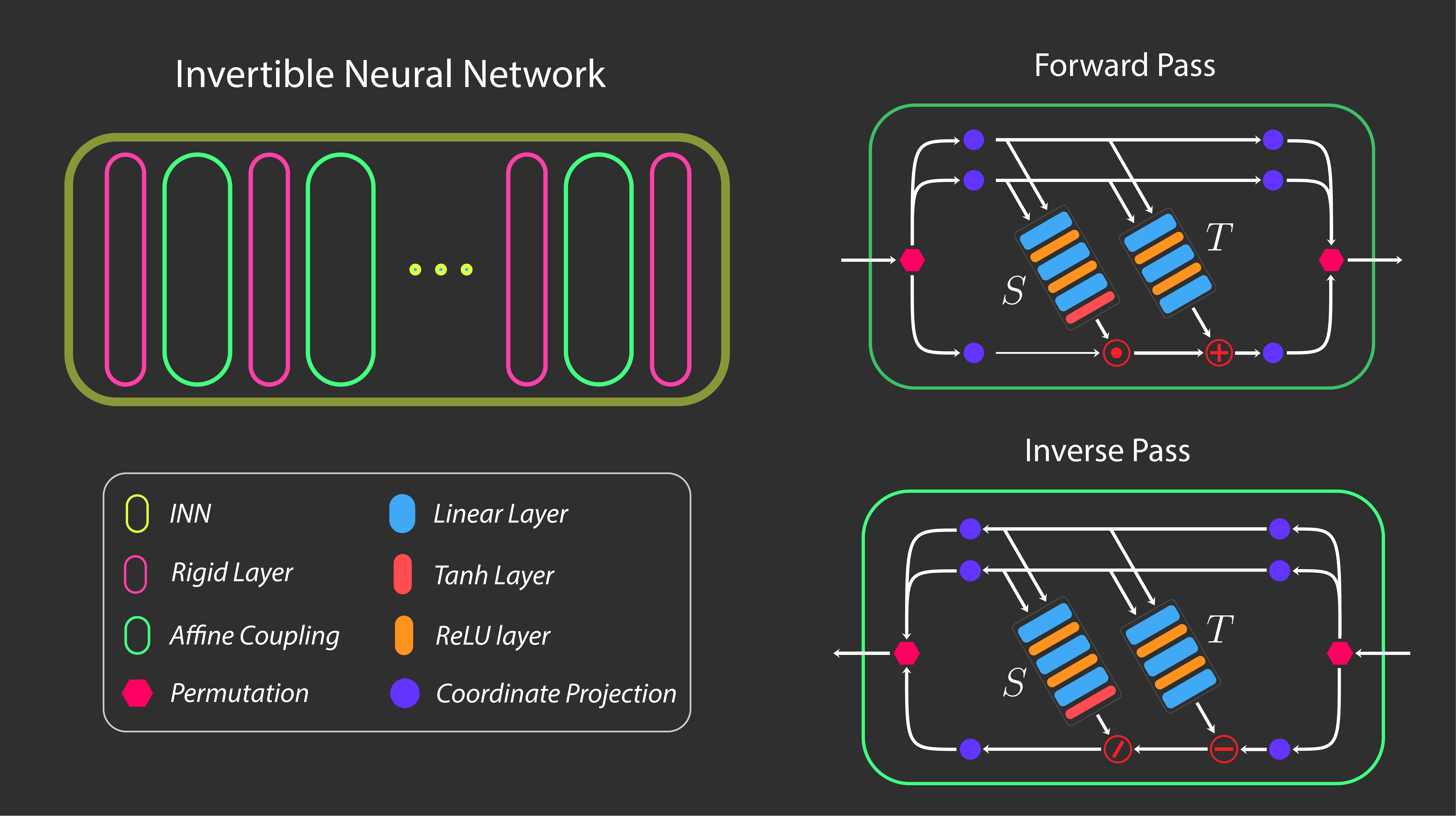}}
\caption{Network Architecture of the INN.}
\label{fig:networkArchitecture}
\end{figure}

\section{Network Architecture}

We represent the family of homeomorphisms $\mathcal{H}_{\phi}$ using an Invertible Neural Network (INN), where $\phi$ denotes the INN parameters. The network architecture of the INN is illustrated in Fig. \ref{fig:networkArchitecture}. The INN consists of an alternating sequence of Rigid Transformation layers and Affine Coupling Layers \cite{dinh2016density}.

\noindent \textbf{Rigid Transformation layer.} This layer consists of a rotation in $\mathbb{R}^3$ along with a translation in $\mathbb{R}^3$, which get updated during each iteration of the optimization process. Both rotation and translation are invertible transformation. The inverse of a rotation is yet another rotation, and the inverse of a translation is yet another translation. This makes the Rigid Transformation layer an invertible layer whose inverse is explicitly available.

\noindent \textbf{Affine Coupling layer.} 
Each Affine Coupling layer consists of two Multi Layer Perceptrons (MLP) : $S$ for scaling and $T$ for translation. It also contains a permutation function $\sigma$. Both $S$ and $T$ take two values as input and output a single value, representing the scaling factor and the translational shift, respectively. During the forward pass, given an input $p = (x,y,z)$, it is first permuted by $\sigma$ to obtain $p' = (u,v,w) = \sigma(p)$. Then $p'$ is modified to get $q'=(u',v',w')$, where $u'=u$, $v'=v$ and $w' = w \cdot S(u,v) + T(u,v)$. Only $w$ is modified using an affine transformation whose scaling and translation factor depend on $u$ and $v$, whereas $u$ and $v$ are left unmodified. The final output $q$ obtained at the end of the forward pass is given by $q = \sigma^{-1}(q')$. During the inverse pass, given an input $p = (x,y,z)$, it is first permuted by $\sigma^{-1}$ to obtain $p' = (u,v,w) = \sigma^{-1}(p)$. Then $p'$ is modified to get $q'=(u',v',w')$, where $u'=u$, $v'=v$ and $w' = \frac{w - T(u,v)}{S(u,v)}$. The final output $q$ obtained at the end of the inverse pass is given by $q = \sigma(q')$.
It can be seen that the inverse pass is indeed the inverse operation of the forward pass, due to the invertibility of the affine function. The permutation $\sigma$ for each Affine Coupling layer is randomly chosen during initialization and is fixed thereafter during the entire optimization process.

The Rigid Transformation layers and Affine Coupling Layers are both invertible layers. Since the composition of invertible functions is invertible, our overall network is invertible. Moreover, each component in our network is continuous, and the composition of continuous functions is continuous, making our overall network continuous. Our network $\mathcal{H}_{\phi}$ is thus, bijective and bicontinuous, and represents a homeomorphism for each value of $\phi$.

\begin{figure}[!t]
\centering
\centerline{\includegraphics[width=1.0\linewidth]{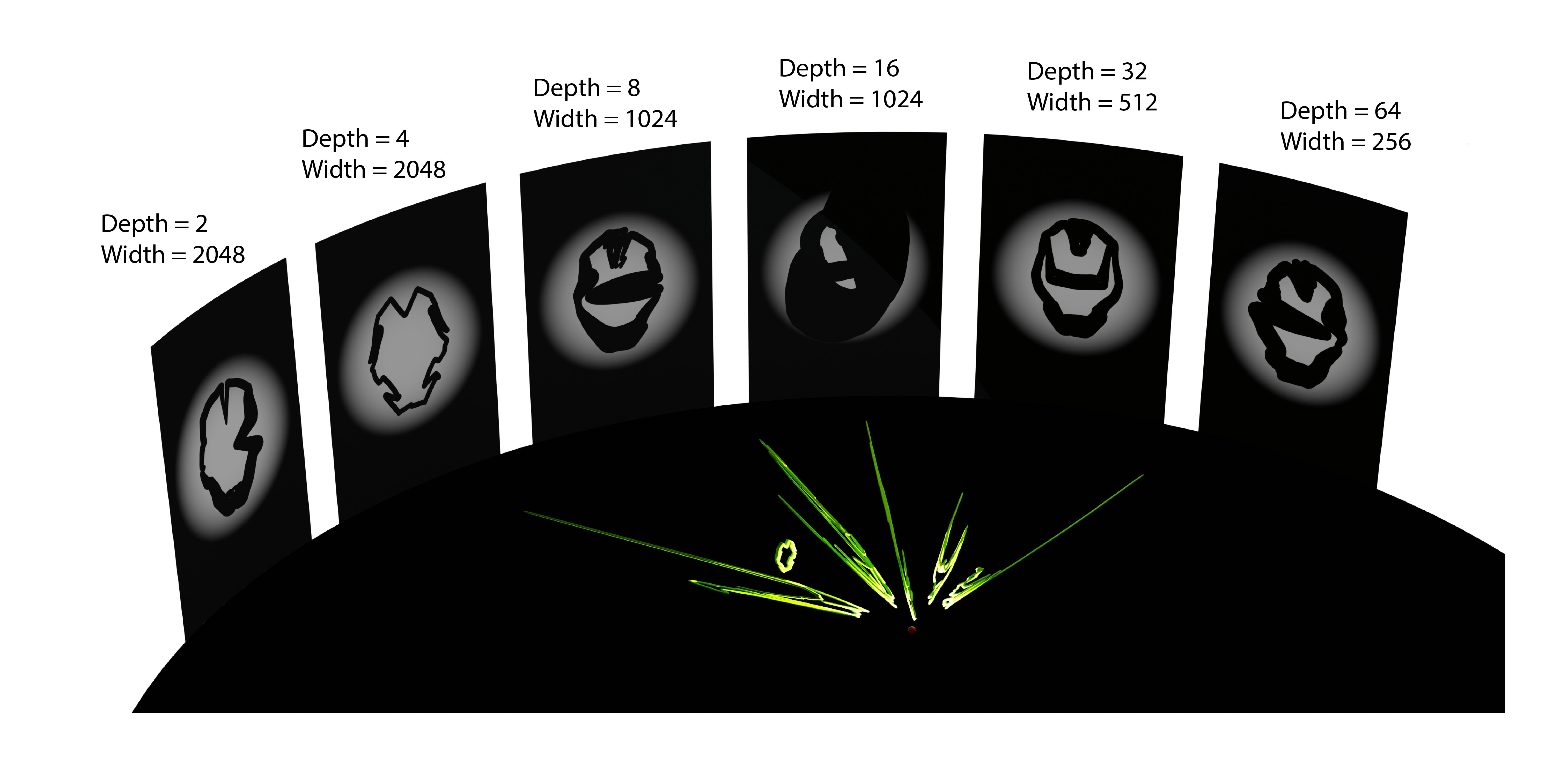}}
\caption{3D Embeddings and Renderings for ablation results on the considered INN architecture for different values of the network depth and width.}
\label{fig:NetworkAblation4}
\end{figure}

\begin{table}[!t]
\centering
\begin{tabular}{|c|c|}
\hline
\textbf{Hyper-parameter}                   & \textbf{Value} \\ \hline
$\theta = (\theta_x, \theta_y, \theta_z)$ & $(0, 0, 0)$             \\ \hline
$t = (t_x, t_y, t_z)$                     & $(0, 0, 0)$      \\ \hline
$f = (f_x, f_y, f_z)$                     & $(1, 1, 2)$      \\ \hline
$(z_-, z_+)$                          & $(10^{-3}, 10^3)$                  \\ \hline
$(W, H)$                                  & $(256, 256)$     \\ \hline
$r$                                       & $0.05$           \\ \hline
$N$                                       & $1000$           \\ \hline
$|\mathcal{B_M}|$     & 100000         \\ \hline
$\tau$     & 100         \\ \hline
\end{tabular}
\caption{Table showing defined hyper-parameters along with their values}
\label{Tab:1}
\end{table}

\begin{table}[]
\centering
\begin{tabular}{|c|c|}
\hline
\textbf{Hyper-parameter} & \textbf{Value} \\ \hline
$w_{img}$               & $1$              \\ \hline
$w_{len}$               & $10^{-2}$              \\ \hline
$w_{mob}$               & $10^{-3}$           \\ \hline
$w_{occ}$               & $10^{-2}$              \\ \hline
$w_{bend}$              & $10^{-2}$              \\ \hline
\end{tabular}
\caption{Table indicating the weights associated with each loss function.}
\label{Tab:2}
\end{table}

\section{Network Ablation}

In Fig. \ref{fig:NetworkAblation4} we show the 3D embeddings and renderings for network ablations on the considered target image. The effect of varying the network depth and width on the expressivity of the network is observed. 

\section{Video Animations}
Videos have been included along with the supplementary which depict the results on the following two scenarios:
\begin{enumerate}
    \item \textbf{Dancer Video}. Given a video of a person dancing as the target input, the spatio-temporal embedding of the knot produced by our approach is displayed in the video from both side view and front view.
    \item \textbf{Evolution of Knot Configuratioan during training}. The evolution of the knot embedding during the optimization process is shown for two target images - Ironman and Bunny.
\end{enumerate}

\section{Hyper-parameters}

Hyper-parameters used during optimization have been given in Table \ref{Tab:1} and \ref{Tab:2}.
The INN architecture used during optimization has $\text{depth} = 8$ (number of Affine Coupling layers) and $\text{width}=1024$ (size of the hidden layer in $S$ and $T$ inside each Affine Coupling layer). The optimization was done using the Adam optimizer with a learning rate of $10^{-5}$.

\section{Computing Infrastructure}

To implement the proposed method, we have used Python-$3.8$ with PyTorch-$1.9$ library. The specification of the processor is Intel $i9-10900X$ ($20$ core) @ $4.50$GHz. We have used a single Nvidia Quadro RTX $5000$ with $16$GB for optimization in our proposed approach.

\end{document}